\documentclass[12pt]{article}
 \usepackage{graphicx}  
\usepackage{amsmath}   
\usepackage[compress]{cite}
\usepackage{amssymb}   
\usepackage{bm} 
\usepackage{dcolumn}
\usepackage{ulem}
\usepackage{color}
\usepackage{mathtools}
\usepackage{mathrsfs}
\usepackage{amsfonts}
\usepackage{varioref}
\usepackage{textcomp}
\usepackage{cancel}
\usepackage{float} 
\usepackage{subfloat}
\usepackage{subcaption}
\usepackage{tcolorbox} 
\usepackage{halloweenmath}
\usepackage{empheq}
\usepackage[active]{srcltx}
\usepackage{tikz}
\usepackage{lmodern}
\usepackage{comment}
\usepackage{makeidx}
\usepackage{graphicx}
\usepackage{braket}
\usetikzlibrary{decorations.pathmorphing}
\tikzset{snake it/.style={decorate, decoration=snake}}
\usepackage{fancybox}
\date{} 
\RequirePackage[colorlinks,citecolor=blue,urlcolor=magenta,linkcolor=blue]{hyperref}
\allowdisplaybreaks
 
\usepackage[top=80pt,bottom=85pt,left=60pt,right=60pt]{geometry}

\def\be{\begin{equation}}
\def\bea{\begin{eqnarray}}
\def\ee{\end{equation}}
\def\eea{\end{eqnarray}}

\numberwithin{equation}{section}

\begin{document}

\begin{titlepage}

\begin{center}

\vskip .5in 
\noindent

{\Large \bf{Integrability and non-integrability for holographic dual of Matrix model and non-Abelian T-dual of AdS$_5\times$S$^5$  }}

\bigskip\medskip

Jitendra Pal${}^{a^\star}$  and  Sourav Roychowdhury${}^{b^\clubsuit}$

\bigskip\medskip
{\small

${}^a${\it{Department of Physics, Indian Institute of
Technology
Roorkee\\
Roorkee
 247667, Uttarakhand, India}}\\[2mm]

${}^b${\it{School of Physical Sciences, Indian Association for Cultivation of Science,\\ Kolkata 700032, West Bengal, India}}\\[2mm]
	
}

\bigskip\medskip

\vskip 0cm 
{\small \tt ${}^\star$jpal1@ph.iitr.ac.in,
${}^\clubsuit$spssrc2727@iacs.res.in}
\vskip .9cm 
     	{\bf Abstract }

\vskip .1in
\end{center}

\noindent
In this paper we study integrability and non-integrability for type-IIA supergravity background dual to deformed plane wave matrix model.
From the bulk perspective, we estimate various chaos indicators that clearly shows chaotic string dynamics in the limit of small value of the parameter $L$ present in the theory. 
On the other hand, the string dynamics exhibits a non-chaotic motion for the large value of the parameter $L$ and therefore presumably an underlying integrable structure.
Our findings reveals that the parameter $L$ in the type-IIA background acts as an interpolation between a non-integrable theory to an integrable theory in dual SCFTs.

\vfill
\eject

\end{titlepage}

\tableofcontents

\section{Introduction and summary } \label{int}
Examining the chaotic behaviour along with the associated non-integrable structure in the context of holographic duality \cite{Maldacena:1997re,Witten:1998qj} has been one of the thrust areas in modern theoretical research
 for the past couple of decades.
 In the context of gauge/string duality, one encounters only a few handful examples of integrable string solutions \cite{Sfetsos:2013wia,Hollowood:2014qma,Delduc:2013qra,Pal:2022hmh,Roychowdhury:2020zer,Filippas:2019puw},
 most of the cases the string dynamics exhibits chaotic motion \cite{PandoZayas:2010xpn,Basu:2011dg,Basu:2011di,Basu:2011fw,Basu:2012ae,PandoZayas:2012ig,Basu:2013uva,Basu:2016zkr,Panigrahi:2016zny,Giataganas:2013dha,Ishii:2021asw,Roychowdhury:2017vdo,Nunez:2018ags,Nunez:2018qcj,Banerjee:2018ifm,Rigatos:2020hlq,Giataganas:2017guj,Filippas:2019ihy,Filippas:2019bht,Rigatos:2020igd}.

In the context of AdS/CFT correspondence \cite{Maldacena:1997re,Witten:1998qj} non-integrable systems play an important role. 
The main idea behind here is to analyse the semi-classical motion of the probe string trajectories that lead to give various chaos indicators. 
These indicators provide us with information whether the associated phase space of the probe string motion allows the {\it{Kolmogorov-Arnold-Moser}} (KAM) tori, characterising the (quasi)-periodic orbits \cite{PandoZayas:2010xpn,Basu:2011dg,Basu:2011di} to be distorted or not. 
At the classical level, identification of these orbits is the primary step towards unveiling an underlying integrable structure associated with phase-space governed by the dynamics of the semi-classical string.

The holographic principle conjectured that the holographic dual of the semi-classical strings is given by a class of single trace operators in the large $N$ limit in boundary QFT. 
Therefore, the above framework provides a conjecture about the integrability or non-integrability of the dual strongly coupled SCFTs in terms of bulk gravity theory. 
The most powerful and explicit way to prove integrability is by constructing of Lax pairs \cite{Bena:2003wd,Arutyunov:2008if,Stefanski:2008ik,Sorokin:2010wn,Zarembo:2010sg}. 
Along this line of research, the classical integrability of the supergravity backgrounds containing an AdS$_5$ and AdS$_4$ factor have been revealed in  \cite{Bena:2003wd,Arutyunov:2008if,Stefanski:2008ik,Sorokin:2010wn,Zarembo:2010sg,Frolov:2005dj}.
Constructing the Lax pair formulation for the two dimensional string sigma model is challenging task. Instead of searching for Lax pairs, one should look for other methods. Kovacic's algorithm \cite{Roychowdhury:2017vdo,Nunez:2018ags,Nunez:2018qcj,K1,B1,K2} provides a popular approach to disproving classical integrability and studied extensively in the context of gauge/string duality \cite{Filippas:2019ihy,Filippas:2019bht,Rigatos:2020igd}. We briefly discuss  Kovacic’s algorithm in the appendix \ref{app D}.

Recently, integrability/non-integrability of marginally deformed 4d $ \mathcal{N}=2 $ SCFTs  \cite{Nunez:2019gbg} has been extensively studied in \cite{Pal:2023qkv}. 
It is shown that the classical string dynamics exhibits integrable structure when the deformation parameter ($\gamma$) is small enough.
On the other hand, for the large values of the parameter  $\gamma$, the string motion turns out to be chaotic and corresponding phase space dynamics becomes non-integrable.

Following the above spirit, in this present literature we focus on studying integrability/non-integrability of type-IIA supergravity solution dual to{\it{ plane wave matrix model}} \cite{Berenstein:2002jq,Lozano:2017ole,Lin:2004kw,Lin:2005nh} with relevant/irrelevant deformation. 
It can be shown that the type-IIA supergravity background we are interested in can be obtained through dimensional reduction of eleven dimensional M-theory backgrounds preserving 16 supercharges and exhibits $SO(6) \times SO(3) \times R$ 
isometries \cite{Lozano:2017ole,Lin:2005nh,Lin:2004nb}. 
We consider the motion of the semi-classical string on this type-IIA supergravity background and workout various chaotic indicators. 
Our findings eventually lead us to conjecture about the integrability/non-integrability of the dual SCFTs. 

In our present work, the equations of motion of the semi-classical string are difficult to solve analytically.
Hence, we keep our entire analysis numerical along with fixing the winding numbers of the string associated with the $U(1)$-isometric directions of the dual type-IIA metric. 
We analyse the Poincar\'{e} sections of corresponding phase space dynamics along with Lyapunov exponents while varying the length scale parameter ($L$) present in the gravity solution. 
We show that the dynamics of the string is controlled by the parameter $L$.

In our analysis we fixed the winding numbers $\{k_1 = k_2 = 1\}$ and studied various Poincar\'{e} sections,  the associated Lyapunov exponents while varying the parameter $L$ from small to large enough. 
We notice that for small values of the parameter, the string motion becomes chaotic.
On the other hand, the phase space of the corresponding string dynamics exhibits an integrable structure when the parameter $L$ is large enough. 
 Our observations clearly imply the fact that the integrable structure persists for the usual non-Abelian T-dual of $AdS_5 \times S^5$ background \cite{Lozano:2017ole} (where non-Abelian T-duality acts on $S^3 \in AdS_5$), which is the case of very large limit of $L $. 
Hence from the bulk perspective, our observations lead that the parameter $L$ acts as an interpolation between a non-integrable theory (corresponds to small $L$) to a integrable theory  (corresponds to $L$ is very large).
We elaborate more on this as we progress in section (\ref{sec3}).

\vskip .1in

The rest of the paper is organised as follows. 
 In section (\ref{sec2}), we briefly review the matrix model along with the dual type-IIA supergravity background. 
 In section (\ref{sec3}), we study the dynamics of the semi-classical string on the type-IIA supergravity background along with the Poincar\'{e} sections and the Lyapunov exponents upon varying the parameter $L$ present in the background.
 Finally, we draw our conclusion and provide some possible physical explanation of our findings in section (\ref{sec4}).

\section{The matrix model and its holographic dual in type-IIA supergravity } \label{sec2}

To begin with, we briefly review the holographic dual in type-IIA supergravity of the{\it{ Plane Wave Matrix Model}} (PWMM) discussed in \cite{Lozano:2017ole,Lin:2004kw,Lin:2005nh,Lin:2004nb,Roychowdhury:2023hvq}. 
The PWMM is a quantum mechanical model, obtained by the modding out the subgroup $S\hat U(2)$ of $\mathcal N=4$ super Yang-Mills theory (SYM) defined on $R \times S^3$. 
This results integrating out the fields in $\mathcal N=4$ SYM Lagrangian under the subgroup $S\hat U(2)$.
The PWMM exhibits $SO(6) \times SO(3) \times R$ symmetry (compactly expressed by the supergroup $SU(2|4)$) and preserves 16 supercharges. 
Each classical vacua of PWMM has one to one correspondence with the partition of $N$ \cite{Lozano:2017ole}. One can have $n_k$ copies of the $k$-dimensional representation of $SU(2)$—of dimension $k = (2j_k +1)$.
Then $N$ can be expressed as

\begin{eqnarray}
	N = \sum_k^T n_k (2j_{k} +1) \ . 
\end{eqnarray}

The holographic dual in eleven dimensional M-theory background corresponds to the above system has been extensively studied in \cite{Lin:2004kw,Lin:2005nh,Lin:2004nb}. 
Upon dimensional reduction along one $U(1)$ isometric direction results in a new class of ten dimensional type-IIA supergravity background that preserves 16 supercharges. 
The metric of resulting type-IIA background takes the form \cite{Lin:2005nh,Lozano:2017ole}
\begin{eqnarray} \label{expl met}
ds_{10,   IIA}^2 = F \bigg[f_1 dt^2 + f_2  \Big(d\sigma^2 + d\eta^2\Big) 
+ f_3 d\Omega_5^2 
+ f_4 d\Omega_2^2 (\chi , \xi) \bigg] \ , 
\end{eqnarray}
where $d\Omega_5^2$ and $d\Omega_5^2(\chi , \xi)$ are metric of the five-sphere and two-sphere with unit radius respectively
and in global coordinates can be expressed as 
\begin{eqnarray} \label{main met 1}
 d\Omega_5^2 (\zeta, \beta, \theta, \phi, \psi) = 4d\Omega_2^2 (\zeta, \beta) + \cos^2 \zeta  d\Omega_3^2 (\theta, \phi, \psi) \ ;  \  d\Omega_2^2(\chi , \xi) = d\chi^2 + \sin^2\chi d\xi^2 \ . 
\end{eqnarray}

The  type-IIA background in \eqref{expl met} is supported by NS-NS two-form and RR one-form and three-form field along with the background dilaton as 
\begin{eqnarray} \label{main back fields}
B_2 &=& f_5 d\Omega_2 (\chi , \xi)  \ , \cr
C_1 &=& f_6 dt \ ;  \  C_3 = f_7 dt \wedge d\Omega_2   \ , \cr
e^{4\Phi} &=& f_8 \ . 
\end{eqnarray}

The warp factors $F(\sigma,\eta)$ and $f_i(\sigma,\eta)$s in the background \eqref{expl met}-\eqref{main back fields} can be expressed in terms of a potential function $V(\sigma,\eta)$ 
\begin{eqnarray} \label{f}
 &&F = \bigg( \frac{2\dot V - \ddot V}{V^{\prime \prime}}\bigg)^{\frac{1}{2}} \ ;  \  f_1 = \frac{4 \ddot V}{2\dot V - \ddot V} \ ;  \  f_2 = - \frac{2V^{\prime \prime}}{\dot V} \ ,  \cr
 &&f_3 = 4 \ ; \  f_4 = \frac{2V^{\prime \prime}\dot V}{\Delta} \ ;  \  f_5 = 2 \bigg(\frac{\dot V \dot V^\prime}{\Delta} + \eta\bigg) \ , \cr
 &&f_6 =  \frac{2\dot V \dot V^\prime}{2\dot V - \ddot V} \ ;  \  f_7 =  -  \frac{4 \dot V^2 V^{\prime \prime}}{\Delta}  \ ;  \  f_8 =   \frac{4 (2\dot V - \ddot V)^3}{V^{\prime \prime} \dot V^2 \Delta^2} \ . 
\end{eqnarray}
The two-dimensional potential function $V(\sigma,\eta)$ satisfies Laplace like equation of the form 
\begin{eqnarray} \label{lap}
 \ddot V + \sigma^2 V^{\prime \prime} = 0  \ . 
\end{eqnarray}
The explicit expressions of the dot and prime of the potential $V(\sigma,\eta)$ along with $\Delta$ is given by 
\begin{eqnarray} \label{dot prime}
&& \dot{V} = \sigma \partial_\sigma V \ ;  \  V^\prime = \partial_\eta V \ ;  \  \ddot{V} =  \sigma \partial_\sigma \dot{V} \ ;    \  V^{\prime\prime} = \partial_\eta^2 V \ ;   \  \dot{V}^\prime =  \sigma \partial_\sigma V^\prime \ , \cr
&&\Delta =  \big(\ddot{V} -2\dot{V} \big) V^{\prime\prime} - \big(\dot{V}^\prime\big)^2 . 
\end{eqnarray}

\section{String motion in type-IIA background } \label{sec3}

We now consider that the bosonic string propagating over the background given in \eqref{expl met} in the presence of the NS-NS two-form \eqref{main back fields}. 
The resulting sigma model could be expressed as
\begin{eqnarray} \label{action}
\mathcal S_p=  - \frac{1}{2} \int d\tau d \tilde \sigma  \bigg[\eta^{\lambda \rho} G_{\mu \nu}  + \epsilon^{\lambda \rho} B_{\mu \nu} \bigg] \partial _\lambda X^\mu  \partial _\rho X^\nu \ . 
 \end{eqnarray}

Here, $\{\lambda, \rho\}$ denote the world-sheet coordinates $(\tau, \tilde \sigma)$ and $\{\mu, \nu\}$ denote the spacetime coordinates.
$G_{\mu \nu}$ is the metric of the background, $B_{\mu \nu}$ is the NS-NS two-form and $X^\mu$ are the string embedding coordinates on the world-sheet. 
Moreover, here $\eta_{\lambda \rho}$ is the two dimensional world-sheet metric of the form $- \eta_{\tau \tau} =  \eta_{ \tilde \sigma \tilde \sigma} = 1$.
Together with above, we fix the convention for the 2d Levi-Civita as, $- \epsilon_{\tau  \tilde \sigma} =  \epsilon_{ \tilde \sigma \tau} = - 1$.

To begin with, we consider that the string sits at $\zeta = \frac{\pi}{2}$ and wraps the isometric directions $\beta$ and $\xi$ of the background \eqref{expl met}. 
Given this fact, we propose a string embedding of the following form
\begin{eqnarray} \label{emb}
&&t = t( \tau ) \ ; \ \sigma = \sigma (\tau) \ ;  \  \eta = \eta(\tau) \ ;  \  \chi = \chi (\tau) \ ,    \cr
 &&  \beta = \beta (\tilde \sigma) = k_1 \tilde \sigma  \ ;   \ \xi = \xi (\tilde \sigma) = k_2 \tilde \sigma  \ . 
\end{eqnarray}

Here, $k_i \ ; \ i =1,2$ are the integers that denote the winding numbers of the string along the isometric directions $\beta$ and $\xi$ respectively. 
Considering the string embedding as proposed in \eqref{emb}, the world-sheet Lagrangian takes the following form 
\begin{eqnarray} \label{lag exp}
\mathcal L_p = \frac{1}{2} F \Bigg[f_1 \dot t^2 + f_2 \big(\dot \sigma^2 + \dot \eta^2\big) - 4f_3 k_1^2 +  f_4 \big(\dot \chi^2 - k_2^2\sin^2\chi\big) \Bigg] + k_2 f_5 \sin \chi \dot \chi \ . 
 \end{eqnarray}
where dot denotes the derivative with respect to the world-sheet time coordinate $\tau$. 

From the Lagrangian \eqref{lag exp}, the Hamiltonian of the system takes the form
\begin{eqnarray} \label{ham}
\mathcal H = \frac{1}{2F} \Bigg[\frac{1}{f_1} p_t^2 + \frac{1}{f_2} \big(p_\sigma^2 + p_ \eta^2\big) + \frac{1}{f_4} \big(p_\chi - k_2 f_5 \sin \chi\big)^2\Bigg] + F \Bigg[\frac{1}{2} f_4 k_2^2 \sin^2\chi + 2 f_3 k_1^2\Bigg] \ .  
 \end{eqnarray}
Using \eqref{ham}, the Hamilton's equations of motion can be read as  
\begin{eqnarray} \label{ham eq}
\dot \sigma &=&  \frac{p_{\sigma} }{F f_2}  \ , \cr
\dot p_{\sigma} &=& -\Big(2k_1^2 f_3 + \frac{1}{2} k_2^2 f_4 \sin^2\chi\Big) \partial_\sigma F +  \Bigg[\frac{E^2}{f_1} + \frac{1}{f_2} \big(p_\sigma^2 + p_ \eta^2\big) + \frac{1}{f_4} \big(p_\chi - k_2 f_5 \sin \chi\big)^2\Bigg] \frac{\partial_\sigma F}{2F^2 }  \cr
&&- \frac{1}{2F}  \Bigg[\frac{\partial_\sigma f_1}{f_1^2} E^2  +  \frac{\partial_\sigma f_2}{f_2^2} \Big(p_\sigma^2 + p_\eta^2\Big) + \frac{\partial_\sigma f_4}{f_4^2} \Big(p_\chi - k_2 f_5 \sin \chi\Big)^2 \cr
&&+ 2k_2 \frac{\partial_\sigma f_5}{f_4}  \sin \chi \Big(p_\chi -k_2 f_5 \sin \chi\Big) \Bigg] 
- F \Bigg[2k_1^2 \partial_\sigma f_3 + \frac{1}{2} k_2^2 \sin^2\chi \partial_\sigma f_4 \Bigg] \ ,  \cr
\dot \chi &=&  \frac{1 }{F f_4} \Big(p_\chi - k_2 f_5 \sin \chi\Big) \ , \cr
\dot p_{\chi} &=& \frac{k_2 f_5 \cos\chi}{F f_4}  \Big(p_\chi - k_2 f_5 \sin \chi\Big) - F f_4  k_2^2 \sin \chi \cos \chi  \ ,    \cr
\dot \eta &=&  \frac{p_{\eta} }{F f_2}  \ , \cr
\dot p_{\eta} &=& -\Big(2k_1^2 f_3 + \frac{1}{2} k_2^2 f_4 \sin^2\chi\Big) \partial_\eta F +   \Bigg[\frac{E^2}{f_1} + \frac{1}{f_2} \big(p_\sigma^2 + p_ \eta^2\big) + \frac{1}{f_4} \big(p_\chi - k_2 f_5 \sin \chi\big)^2\Bigg] \frac{\partial_\eta F}{2F^2 }\cr
&&- \frac{1}{2F}  \Bigg[\frac{\partial_\eta f_1}{f_1^2} E^2  +  \frac{\partial_\eta f_2}{f_2^2} \Big(p_\sigma^2 + p_\eta^2\Big) + \frac{\partial_\eta f_4}{f_4^2} \Big(p_\chi - k_2 f_5 \sin \chi\Big)^2 \cr
&&+ 2k_2 \frac{\partial_\eta f_5}{f_4}  \sin \chi \Big(p_\chi -k_2 f_5 \sin \chi\Big) \Bigg]  - F \Bigg[2k_1^2 \partial_\eta f_3 + \frac{1}{2} k_2^2 \sin^2\chi \partial_\eta f_4 \Bigg] \ . 
\end{eqnarray}
The above equations of motion are supplemented by the Virasoro constraints $T_{\tau \tau} = 0 = T_{ \tau \tilde \sigma}$,
where $T_{\lambda \rho} \ ;  \  \lambda , \rho = \{\tau, \tilde \sigma\}$ is the two-dimensional world-sheet stress tensor 
 with the expressions\footnote{It is trivial to show that the string embedding given in \eqref{emb}, the $T_{\tau \tilde \sigma}$ component of the 2d stress tensor vanishes identically. } 
 \begin{eqnarray} \label{vira expl}
T_{\tau \tau} = \frac{1}{2} G_{\mu \nu} \Big(\partial_\tau X^\mu \partial_\tau X^\nu + \partial_{\tilde \sigma} X^\mu \partial_{\tilde \sigma} X^\nu\Big)  = 0  \  ;  \  T_{\tau \tilde\sigma} =  \frac{1}{2} G_{\mu \nu} \partial_\tau X^\mu \partial_{\tilde \sigma}  X^\nu = 0  \ . 
 \end{eqnarray}

Moreover, it can be shown that the Hamiltonian in \eqref{ham} can be read as time-time component of the world-sheet stress tensor $\mathcal H = T_{\tau \tau}$.

Due to the complicated warp factors (see \ref{app w}), it is very difficult to carry out the solution of equations of motion of the probe string analytically as well as to study Kovacic’s algorithms.
Due to this, we keep our entire analysis numerical. We now numerically estimate the Poincar\'{e} sections by solving the Hamilton's equations of motion \eqref{ham eq} for the coarse-grain potential $V(\sigma, \eta)$ associated to the deformed PWMM solution subjected to the Virasoro constraints $T_{\tau \tau} = \mathcal H = 0$. 
The explicit expression of the potential $V (\sigma, \eta)$ is given by\footnote{see Appendix (\ref{app w}) for the explicit expressions of the corresponding warp factors present in type-IIA solution as given in \eqref{f}. } \cite{Lozano:2017ole,Roychowdhury:2023hvq}
\begin{eqnarray} \label{pot}
 V(\sigma , \eta) &=&    \eta \Big(\sigma^2 - \frac{2}{3} \eta^2\Big) - \alpha  \Bigg[2\eta \ln\sigma - \sqrt{\sigma^2 + \big(L-\eta\big)^2} +  \sqrt{\sigma^2 + \big(L+\eta\big)^2} \cr
 && -  \eta  \ln \bigg[\Big( L-\eta + \sqrt{\sigma^2 + \big(L-\eta\big)^2}\Big) \Big(L+\eta + \sqrt{\sigma^2 + \big(L+\eta\big)^2}\Big) \bigg] \Bigg] \  , 
 \end{eqnarray} 
where the $\eta$-coordinate is bounded between $0$ to $L$, $\eta \in [0,L]$. 
It is shown that for large $L$ limit the potential in \eqref{pot} corresponds to  the potential that provides the non-Abelian T-dual of AdS background (namely non-Abelian T-dual of $AdS_5 \times S^5$ ; where non-Abelian T-duality acts on $S^3$ inside the $AdS_5$ subspace) and can be interpreted 
as the {\it{deformation}} of the dual  PWMM by an irrelevant operator \cite{Lozano:2017ole}. 
On the other hand, for $\eta = L$ the background generated by the corresponding potential in \eqref{pot} can be interpreted as a deformation of PWMM by relevant operator. 

\subsection{Poincar\'{e} section}

The study of phase space trajectories in dynamical systems is closely related to the phenomenon of chaos. In chaotic systems, trajectories strongly depend on initial conditions, indicating that even slight changes in the initial state can lead to substantially different trajectories over time.  
The unpredictability and aperiodic motion of trajectories in phase space frequently illustrate this sensitivity to initial conditions.
In contrast, in integrable systems, trajectories are confined to invariant tori and exhibit regular, predictable behaviour. Hence, the study of dynamical systems phase space trajectories plays a crucial role in probing the system's underlying (non)-integrable structure.

 The KAM theorem \cite{p1,p2}, typically formulated in the context of trajectories in phase space, addresses the behaviour of an integrable Hamiltonian system. The KAM theorem states that under certain conditions, small perturbations of an integrable Hamiltonian system will not significantly disrupt the regular behaviour of its trajectories.
 In such systems, the motion is typically confined to invariant tori, and each invariant tori corresponds to a specific set of initial conditions in the system. These tori are $N$-dimensional hypersurfaces in $2N$-dimensional phase space where the trajectories of the system remain confined, reflecting the regular behaviour of the system's dynamics.

To examine the integrable structure of the phase space in a dynamical system, one can slice the N-dimensional foliated KAM tori using a lower-dimensional hypersurface. By doing so, we can determine whether many foliated KAM curves remain intact. These cross-sections of the foliated KAM tori are called Poincaré sections. As chaos emerges, the orderly structure of this foliation is disrupted, leading to the breakdown of its organized pattern. Eventually, we observe a random distribution of phase points indicative of the chaotic behaviour of the system.

\begin{figure}[ H]
	\centering
	\hfill
	\begin{subfigure}{0.35\textwidth}
		\includegraphics[width=\textwidth]{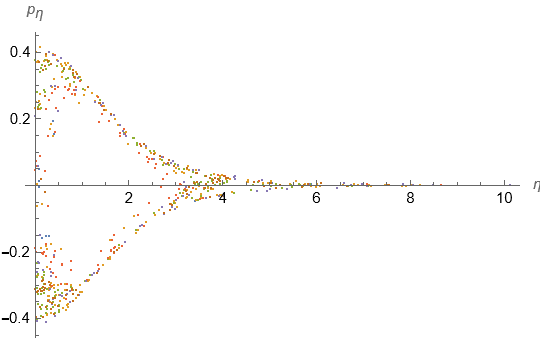}
		\caption{$E=20$, $\alpha=0.1$, $L=1$.  Initial data: $\sigma (0) = 2 ,  \ p_\sigma (0) = 0 $ . }
		\label{fig:1}
	\end{subfigure}
	\hfill
	\begin{subfigure}{0.35\textwidth}
		\includegraphics[width=\textwidth]{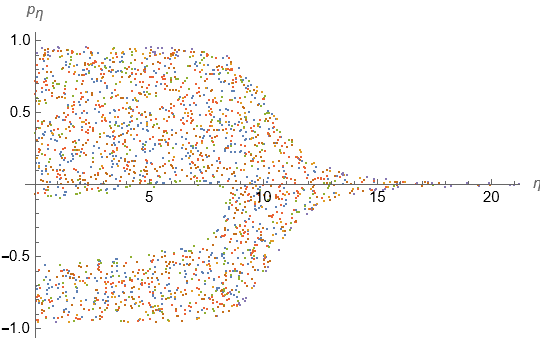}
		\caption{$E=20$, $\alpha=0.1$, $L=10$.  Initial data: $\sigma (0) = 2 ,  \ p_\sigma (0) = 0 $ . }
		\label{fig:2}
	\end{subfigure}
	\hfill
	\begin{subfigure}{0.35\textwidth}
		\includegraphics[width=\textwidth]{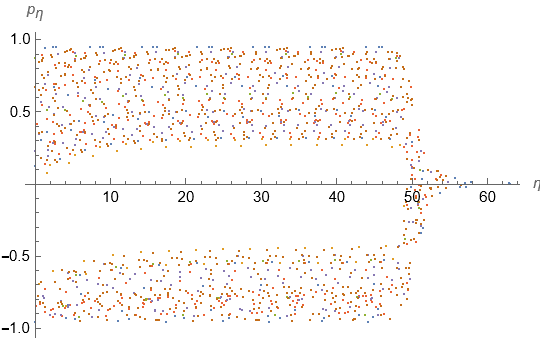}
		\caption{$E=20$, $\alpha=0.1$, $L=50$.  Initial data: $\sigma (0) = 2 ,  \ p_\sigma (0) = 0 $ . }
		\label{fig:3}
	\end{subfigure}
	\hfill
       \begin{subfigure}{0.35\textwidth}
	\includegraphics[width=\textwidth]{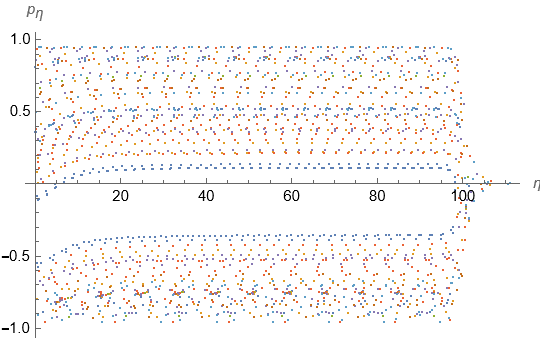}
	\caption{$E=20$, $\alpha=0.1$, $L=100$.  Initial data: $\sigma (0) = 2 ,  \ p_\sigma (0) = 0 $ . }
	\label{fig:2}
        \end{subfigure}
	\hfill
       \begin{subfigure}{0.35\textwidth}
	\includegraphics[width=\textwidth]{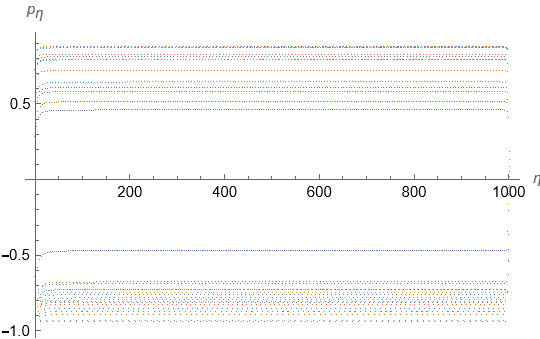}
	\caption{$E=20$, $\alpha=0.1$, $L=1000$.  Initial data: $\sigma (0) = 2 ,  \ p_\sigma (0) = 0 $ .  }
	\label{fig:4}
        \end{subfigure}
        \hfill
\begin{subfigure}{0.35\textwidth}
	\includegraphics[width=\textwidth]{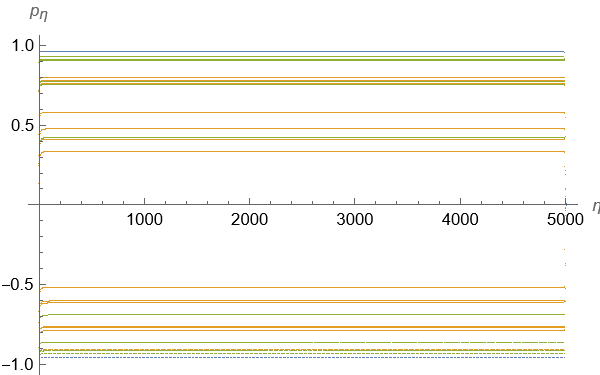}
	\caption{$E=20$, $\alpha=0.1$, $L=5000$.  Initial data: $\sigma (0) = 2 ,  \ p_\sigma (0) = 0 $ . }
	\label{fig:5}
\end{subfigure}
	\hfill
\begin{subfigure}{0.35\textwidth}
	\includegraphics[width=\textwidth]{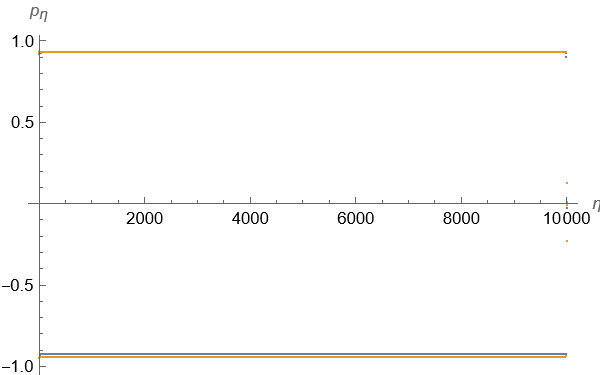}
	\caption{$E=20$, $\alpha=0.1$, $L=10000$.  Initial data: $\sigma (0) = 2 ,  \ p_\sigma (0) = 0 $ . }
	\label{fig:6}
\end{subfigure}
\caption{ Poincar\'{e} sections for the semi-classical string motion in type-IIA background. 
        In the plots we fix the energy of the string as
       $E=20$ and varying the parameter
       $L$ from small to very large. 
       }
       \end{figure}

In our phase space plot we fixed the energy of the string to be $E=20$ and the parameter $\alpha$ as  $\alpha= 0.1$. 
Moreover, we varying the parameter $L$ starting from a small value to large : $\{L\}=\{1, 10, 50, 100, 1000, 5000, 10000\}$. 
The corresponding phase space plots are given in Figs.(\ref{fig:1})-(\ref{fig:6}). 
In the plots we set the winding numbers of the strings $k_1= k_2 =1$ (for winding numbers $\{k_1 , k_2\} > 1$, see Appendix (\ref{app A}) for details).

We now consider different sets of initial conditions that generate solutions to the Hamilton equation given in \eqref{ham eq} for different values of the parameter $L$ as described previously.
 The initial conditions are chosen such that they satisfy the Virasoro constraint {\it{i.e.}}  $\mathcal H = 0$.
 In the present analysis, we consider the initial conditions as $\sigma(0)=2=$ const. and $p_{\sigma}(0)=0$. 
Moreover, in our analysis the four dimensional phase space is characterised by generalised coordinates and momenta labelled by: $\{\chi , p_\chi , \eta, p_\eta\}$. For the small value of the parameter $L$, the orderly structure of KAM tori gets destroyed in our system \cite{PandoZayas:2010xpn,Basu:2011dg,Basu:2011di}, leading to the breakdown of its organized pattern. Eventually, we see a scattered arrangement of phase space points, which is exhibited by a non-integrable system. It turns out that the nature of the plots do not significantly change for other small as well as the large values of $L$ that we presented below. Therefore, the main conclusion will remain unchange even for those small as well large values of $L$.

Further, we also consider the $L = \infty$ of the potential given in (\ref{pot}). This is the special limit called the non-Abelian
T-dual solution. This background preserves integrability. We have given the plots of the Poincar\'{e} section for the non-Abelian
T-dual background in Appendix C. Here, we get the phase space dynamics exhibits a non-chaotic behaviour indicating the integrable structure. 

\subsection{ Lyapunov exponents}

In the following we numerically compute the {\it{Lyapunov exponent}} \cite{PandoZayas:2010xpn,Basu:2011dg,Basu:2011di},
which is another chaos indicator. 
For a chaotic system, during the time evolution the system becomes sensitive depending on the choice of initial conditions imposed on the dynamics. 
The Lyapunov exponent ($\lambda$) provides the deviation between two nearby trajectories in the phase space dynamics due to a variation (small) of the initial conditions \footnote{The expression of the Lyapunov exponent is given by  \cite{PandoZayas:2010xpn,Basu:2011dg,Basu:2011di}
\begin{equation}\label{Def:Lyap}
\lambda = \lim_{\tau\rightarrow\infty}\lim_{\Delta X_{0}\rightarrow\infty}
\frac{1}{\tau}\ln \bigg[\frac{\Delta X\big(X_{0},\tau\big)}{\Delta X\big(X_{0},0\big)}\bigg] \ ,
\end{equation}
here $\Delta X\big(X_{0},\tau\big)$ measures the separation between the infinitesimally close trajectories in the phase space after sufficiently late times.}. These are the signature trademark of the chaotic motion and
encode the sensitivity of the phase space trajectories on the initial conditions. For a chaotic system $\lambda$
converges to a positive value whereas, $\lambda$ decays to zero for a non-chaotic (non-dissipative) dynamical system.

\begin{figure}[ H]
	\centering

	\begin{subfigure}{0.35\textwidth}
		\includegraphics[width=\textwidth]{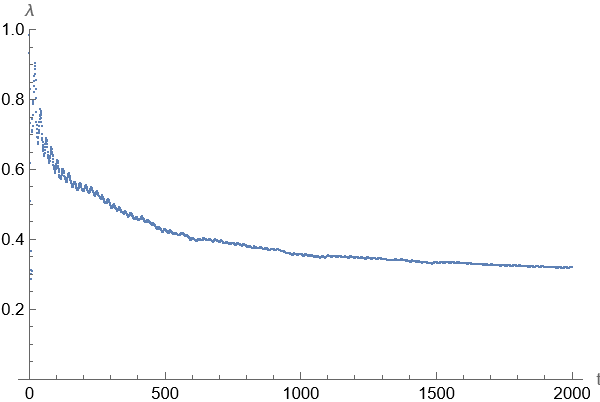}
		\caption{$L=10$ }
		\label{fig:3}
	\end{subfigure}
	\hfill
	\begin{subfigure}{0.35\textwidth}
		\includegraphics[width=\textwidth]{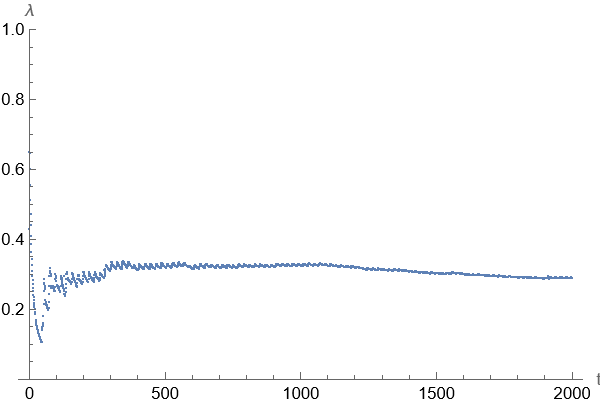}
		\caption{$L=50$ }
		\label{fig:2}
	\end{subfigure}
	\hfill
	\begin{subfigure}{0.35\textwidth}
		\includegraphics[width=\textwidth]{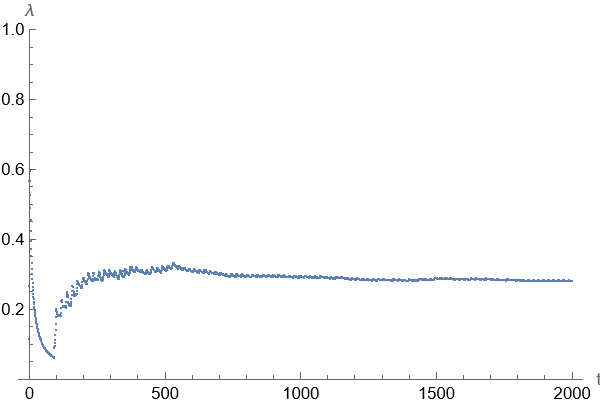}
		\caption{$L=100$ }
		\label{fig:4}
	\end{subfigure}
	\hfill
	\begin{subfigure}{0.35\textwidth}
		\includegraphics[width=\textwidth]{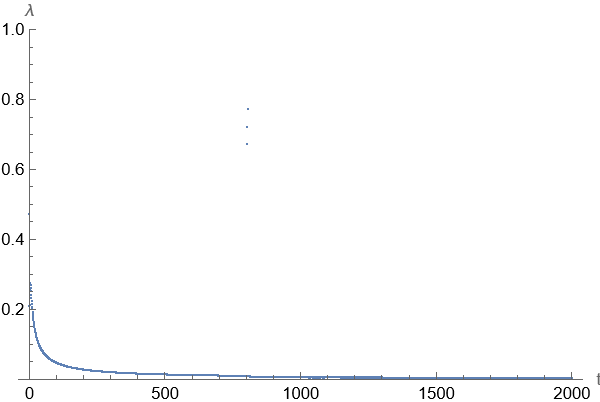}
		\caption{$L=5000$ }
		\label{fig:5}
	\end{subfigure}

	\caption{ Plots of Lyapunov exponent(s) for
		the motion of the semi-classical string. 
		We consider different values of parameter $L: \{10, 50, 100, 5000\}$ as indicated in the figure while keeping the energy of the string fix at $E=20$ and $\alpha = 0.1$.}
\end{figure}

In Fig.(2) we compute the Lyapunov exponent of our system. We consider the energy of the string as $E=20$ along with the initial separation $\Delta X (0) = 10^{-7}$ (cf. \eqref{Def:Lyap}).
 In Fig.(2), we observe that upon increasing the values of the parameter $L$, the Lyapunov exponent ($\lambda$) shows a sharp fall towards
 zero for late time \cite{Nunez:2018qcj} indicating corresponding integrable dynamics of the associated phase space.

\section{Summary and outlook  } \label{sec4}

In this work, we studied chaotic dynamics of a semi-classical string moving in a class of type-IIA supergravity background that preserves 16 supercharges. 
The holographic dual of this background is characterised by relevant/irrelevant deformation of the plane wave matrix model depends on the parameter $L$ present in the background  \cite{Lozano:2017ole}. 
Small $L$ corresponds to relevant deformation and large $L$ corresponds to irrelevant deformation of PWMM.

The main aim of this work was to examine the dynamics of the semi-classical string upon varying the parameter $L$ in a wide range. 
In the core part of our analysis we consider the Hamiltonian framework and examine Poincar\'{e} sections along with the Lyapunov exponents ($\lambda$) for different values of the parameter $L$.

In our analysis we obtain distorted KAM tori as well positive Lyapunov exponents for small values of the parameter $L$ present in the type-IIA supergravity background. 
This leads us to interpret that the string motion exhibits chaotic behaviour and the system is non-integrable for the small value of the parameter $L$. 
On the other hand, we observed that string dynamics exhibits an integrable structure for the large value of the parameter $L$ along with vanishing Lyapunov exponent ($\lambda |_{L>>1} \sim 0$).
Therefore, our findings show that the parameter $L$ in the dual type-IIA background acts as an interpolation between a non-integrable theory to an integrable theory in dual SCFTs.

From the bulk perspective, the non-integrability of the string dynamics could emerge from the corresponding electrostatic configurations discussed in \cite{Lin:2005nh}.
In the corresponding electrostatic configuration, addition to the infinite conducting plate at $\eta=0$ we have a set of finite size conducting disks with finite charge $Q_i$ at regular intervals along $\eta$-direction ($\eta_i$).
The presence of these finite size conducting disks along the holographic direction $\eta$ could be an artefact for chaotic string dynamics we observed for small values of the parameter $L$. 
On the other hand, for the very large values of the parameter $L$, we have only the infinite conducting plate at $\eta=0$ and the finite size conducting disks move to very far distance along $\eta$-direction and that makes string motion smooth and integrable. 
However,  the precise reason of the interpolating string dynamics is lacking in this present literature and requires further investigation. 
Finally, it will be interesting to identify the precise connection between our findings and the works revealed in \cite{Asano:2015eha} and find out the origins of this interpolating behaviour in the dual gauge theory perspective.  
We hope to report along this line of investigation in the near future.

\vskip .3in
 
\noindent {\bf\large Acknowledgement}

 \vskip .1in

\noindent
We are  indebted to Dibakar Roychowdhury for his insights and discussion on several parts of our work. 
We would also like to express our gratitude to Carlos Nunez for clarifying several issues along with his key insights on various parts of our work.


\appendix

\section{Warp factors in type-IIA solution } \label{app w}

For the potential given in \eqref{pot}, the corresponding $F$ and $f_i$s in the type-IIA supergravity solution \eqref{expl met}-\eqref{main back fields} take the form  
\begin{eqnarray}
	F=	\frac{\sqrt{F_N}}{\sqrt{4 \eta -\alpha 
			\bigg(\frac{1}{\sqrt{(\eta +L)^2+\sigma ^2}}-\frac{1}{\sqrt{(L-\eta )^2+\sigma ^2}}\bigg)-\frac{\alpha  L (\eta
				-L)}{\big((L-\eta )^2+\sigma ^2\big)^{3/2}}-\frac{\alpha  L (\eta +L)}{\big((\eta +L)^2+\sigma
				^2\big)^{3/2}}}} \ , 
\end{eqnarray}
\begin{eqnarray*}
	F_N&=& 	 \sqrt{\alpha }\Bigg(\eta ^2 \bigg(\sigma ^2 \bigg(\frac{1}{\big((\eta +L)^2+\sigma
		^2\big)^{3/2}}-\frac{1}{\big((L-\eta )^2+\sigma ^2\big)^{3/2}}\bigg)-\frac{2}{\sqrt{(L-\eta )^2+\sigma
			^2}}+\frac{2}{\sqrt{(\eta +L)^2+\sigma ^2}}\bigg)\\&&+ \ \eta  L \bigg(\sigma ^2 \big(\frac{1}{\big((\eta +L)^2+\sigma
		^2\big)^{3/2}}+\frac{1}{\big((L-\eta )^2+\sigma ^2\big)^{3/2}}\big)+2 \big(\frac{1}{\sqrt{(\eta
			+L)^2+\sigma ^2}}+\frac{1}{\sqrt{(L-\eta )^2+\sigma ^2}}\big)\bigg)\\&&+ \ \sigma ^4 \bigg(\frac{1}{\big((\eta
		+L)^2+\sigma ^2\big)^{3/2}}-\frac{1}{\big((L-\eta )^2+\sigma ^2\big)^{3/2}}\bigg)\Bigg) \ , 
\end{eqnarray*}

	\begin{eqnarray*}
	f_1=-\frac{f_{1N}}{	f_{1D}} \ , 
\end{eqnarray*}

\begin{eqnarray*}
	f_{1N}&=&4 \sigma  \bigg(\alpha  \sigma  \bigg(\frac{\eta ^2+2 L^2-2 \eta  L+\sigma ^2}{\big((L-\eta )^2+\sigma
		^2\big)^{3/2}}+\frac{\sigma ^2}{\big((\eta +L)^2+\sigma ^2\big)^{3/2}}-\frac{2}{\sqrt{(\eta +L)^2+\sigma
			^2}}\bigg)+\frac{\alpha  \eta ^2 \sigma }{\big((\eta +L)^2+\sigma ^2\big)^{3/2}}\\\nonumber&&+ \ \eta  \sigma  \bigg(\alpha  L
	\big(\frac{1}{\big((\eta +L)^2+\sigma ^2\big)^{3/2}}-\frac{1}{\big((L-\eta )^2+\sigma
		^2\big)^{3/2}}\big)+4\bigg)\bigg) \ , 
\end{eqnarray*}

\begin{eqnarray*}
	f_{1D}&=&\alpha  \bigg(\eta ^2 \bigg(\sigma ^2 \big(\frac{1}{\big((\eta
		+L)^2+\sigma ^2\big)^{3/2}}-\frac{1}{\big((L-\eta )^2+\sigma ^2\big)^{3/2}}\big)-\frac{2}{\sqrt{(L-\eta
			)^2+\sigma ^2}}+\frac{2}{\sqrt{(\eta +L)^2+\sigma ^2}}\bigg)\\\nonumber&&+ \ \eta  L \bigg(\sigma ^2 \big(\frac{1}{\big((\eta
		+L)^2+\sigma ^2\big)^{3/2}}+\frac{1}{\big((L-\eta )^2+\sigma ^2\big)^{3/2}}\big)+2
	\big(\frac{1}{\sqrt{(\eta +L)^2+\sigma ^2}}+\frac{1}{\sqrt{(L-\eta )^2+\sigma ^2}}\big)\bigg)\\\nonumber&&+ \ \sigma ^4
	\big(\frac{1}{\big((\eta +L)^2+\sigma ^2\big)^{3/2}}-\frac{1}{\big((L-\eta )^2+\sigma
		^2\big)^{3/2}}\big)\bigg) \ , 
\end{eqnarray*}

\begin{eqnarray*}
	f_2=	-\frac{2 \left(-4 \eta +\alpha  \left(\frac{1}{\sqrt{(\eta +L)^2+\sigma ^2}}-\frac{1}{\sqrt{(L-\eta )^2+\sigma
				^2}}\right)+\frac{\alpha  L (\eta -L)}{\left((L-\eta )^2+\sigma ^2\right)^{3/2}}+\frac{\alpha  L (\eta
			+L)}{\left((\eta +L)^2+\sigma ^2\right)^{3/2}}\right)}{-\frac{\alpha  L^2}{\sqrt{(L-\eta )^2+\sigma
				^2}}-\frac{\alpha  \left(\eta ^2+2 \eta  L+\sigma ^2\right)}{\sqrt{(\eta +L)^2+\sigma ^2}}+\eta  \left(\alpha  L
		\left(\frac{1}{\sqrt{(\eta +L)^2+\sigma ^2}}+\frac{1}{\sqrt{(L-\eta )^2+\sigma ^2}}\right)+2 \sigma
		^2\right)+\alpha  \sqrt{(L-\eta )^2+\sigma ^2}} \ , 
\end{eqnarray*}

\begin{eqnarray*}
	f_3=4 \ , 
\end{eqnarray*}

\begin{eqnarray*}
	f_4=\frac{f_{4N}}{f_{4D}} \ , 
\end{eqnarray*}

\begin{eqnarray*}
	f_{4N}&=&	2 \bigg(-4 \eta +\alpha  \big(\frac{1}{\sqrt{(\eta +L)^2+\sigma ^2}}-\frac{1}{\sqrt{(L-\eta )^2+\sigma
			^2}}\big)+\frac{\alpha  L (\eta -L)}{\big((L-\eta )^2+\sigma ^2\big)^{3/2}}+\frac{\alpha  L (\eta
		+L)}{\big((\eta +L)^2+\sigma ^2\big)^{3/2}}\bigg) \\&& \bigg(-\frac{\alpha  L^2}{\sqrt{(L-\eta )^2+\sigma
			^2}}-\frac{\alpha  \big(\eta ^2+2 \eta  L+\sigma ^2\big)}{\sqrt{(\eta +L)^2+\sigma ^2}}+\eta  \bigg(\alpha  L
	\bigg(\frac{1}{\sqrt{(\eta +L)^2+\sigma ^2}}+\frac{1}{\sqrt{(L-\eta )^2+\sigma ^2}}\bigg)+2 \sigma
	^2\bigg)\\&&+ \ \alpha  \sqrt{(L-\eta )^2+\sigma ^2}\bigg) \ , 
\end{eqnarray*}

\begin{eqnarray*}
	f_{4D}&=&	\bigg(-4 \eta +\alpha  \big(\frac{1}{\sqrt{(\eta
			+L)^2+\sigma ^2}}-\frac{1}{\sqrt{(L-\eta )^2+\sigma ^2}}\big)+\frac{\alpha  L (\eta -L)}{\big((L-\eta
		)^2+\sigma ^2\big)^{3/2}}+\frac{\alpha  L (\eta +L)}{\big((\eta +L)^2+\sigma ^2\big)^{3/2}}\bigg) \\&&
	\bigg(\sigma  \bigg(\alpha  \sigma  \bigg(\frac{\eta ^2+2 L^2-2 \eta  L+\sigma ^2}{\big((L-\eta )^2+\sigma
		^2\big)^{3/2}}+\frac{\sigma ^2}{\big((\eta +L)^2+\sigma ^2\big)^{3/2}}-\frac{2}{\sqrt{(\eta +L)^2+\sigma
			^2}}\bigg)+\frac{\alpha  \eta ^2 \sigma }{\big((\eta +L)^2+\sigma ^2\big)^{3/2}}\\&& + \ \eta  \sigma  \big(\alpha  L
	\bigg(\frac{1}{\big((\eta +L)^2+\sigma ^2\big)^{3/2}}-\frac{1}{\big((L-\eta )^2+\sigma
		^2\big)^{3/2}}\bigg)+4\big)\bigg)+\frac{2 \alpha  L^2}{\sqrt{(L-\eta )^2+\sigma ^2}}\\&&  + \ \frac{2 \alpha 
		\big(\eta ^2+2 \eta  L+\sigma ^2\big)}{\sqrt{(\eta +L)^2+\sigma ^2}}-2 \eta  \big(\alpha  L
	\bigg(\frac{1}{\sqrt{(\eta +L)^2+\sigma ^2}}+\frac{1}{\sqrt{(L-\eta )^2+\sigma ^2}}\bigg)+2 \sigma ^2\big)\\&&- \ 2
	\alpha  \sqrt{(L-\eta )^2+\sigma ^2}\bigg)-\bigg(\sigma ^2 \bigg(\alpha  L \bigg(\frac{1}{\big((\eta
		+L)^2+\sigma ^2\big)^{3/2}}+\frac{1}{\big((L-\eta )^2+\sigma ^2\big)^{3/2}}\bigg)\\&&+ \ \frac{\alpha
	}{\sqrt{(L-\eta )^2+\sigma ^2} \big(-\eta +\sqrt{(L-\eta )^2+\sigma ^2}+L\big)}+\frac{\alpha }{\sqrt{(\eta
			+L)^2+\sigma ^2} \big(\eta +\sqrt{(\eta +L)^2+\sigma ^2}+L\big)} \\&&+ \ 2\bigg)-2 \alpha \bigg)^2 \ , 
\end{eqnarray*}

\begin{eqnarray*}
	f_5=	2 \bigg(\eta +\frac{f_{5N}}{f_{5D}}\bigg) \ , 
\end{eqnarray*}

\begin{eqnarray*}
	f_{5N}&=&\bigg(-\frac{\alpha  L^2}{\sqrt{(L-\eta )^2+\sigma ^2}}-\frac{\alpha  \big(\eta ^2+2 \eta 
		L+\sigma ^2\big)}{\sqrt{(\eta +L)^2+\sigma ^2}}+\eta  \bigg(\alpha  L \bigg(\frac{1}{\sqrt{(\eta +L)^2+\sigma
			^2}}+\frac{1}{\sqrt{(L-\eta )^2+\sigma ^2}}\bigg)+2 \sigma ^2\bigg)\\&&+ \ \alpha  \sqrt{(L-\eta )^2+\sigma ^2}\bigg)\times
	\bigg(\sigma ^2 \big(\alpha  L \bigg(\frac{1}{\big((\eta +L)^2+\sigma ^2\big)^{3/2}}+\frac{1}{\big((L-\eta
		)^2+\sigma ^2\big)^{3/2}}\bigg)\\&&+ \ \frac{\alpha }{\sqrt{(L-\eta )^2+\sigma ^2} \big(-\eta +\sqrt{(L-\eta
			)^2+\sigma ^2}+L\big)}+\frac{\alpha }{\sqrt{(\eta +L)^2+\sigma ^2} \bigg(\eta +\sqrt{(\eta +L)^2+\sigma
			^2}+L\bigg)}\\&&+ \ 2\big)-2 \alpha \bigg) \ , 
\end{eqnarray*}

\begin{eqnarray*}
	f_{5D}&=&\bigg(-4 \eta +\alpha  \big(\frac{1}{\sqrt{(\eta +L)^2+\sigma
			^2}}-\frac{1}{\sqrt{(L-\eta )^2+\sigma ^2}}\big)+\frac{\alpha  L (\eta -L)}{\big((L-\eta )^2+\sigma
		^2\big)^{3/2}}+\frac{\alpha  L (\eta +L)}{\big((\eta +L)^2+\sigma ^2\big)^{3/2}}\bigg) \\&& \bigg(\sigma 
	\bigg(\alpha  \sigma  \bigg(\frac{\eta ^2+2 L^2-2 \eta  L+\sigma ^2}{\big((L-\eta )^2+\sigma
		^2\big)^{3/2}}+\frac{\sigma ^2}{\big((\eta +L)^2+\sigma ^2\big)^{3/2}}-\frac{2}{\sqrt{(\eta +L)^2+\sigma
			^2}}\bigg)+\frac{\alpha  \eta ^2 \sigma }{\big((\eta +L)^2+\sigma ^2\big)^{3/2}}\\&&+ \ \eta  \sigma  \bigg(\alpha  L
	\bigg(\frac{1}{\big((\eta +L)^2+\sigma ^2\big)^{3/2}}-\frac{1}{\big((L-\eta )^2+\sigma
		^2\big)^{3/2}}\bigg)+4\bigg)\bigg)+\frac{2 \alpha  L^2}{\sqrt{(L-\eta )^2+\sigma ^2}}\\&& + \ \frac{2 \alpha 
		\big(\eta ^2+2 \eta  L+\sigma ^2\big)}{\sqrt{(\eta +L)^2+\sigma ^2}}-2 \eta  \bigg(\alpha  L
	\bigg(\frac{1}{\sqrt{(\eta +L)^2+\sigma ^2}}+\frac{1}{\sqrt{(L-\eta )^2+\sigma ^2}}\bigg)+2 \sigma ^2\bigg)\\&& - \ 2
	\alpha  \sqrt{(L-\eta )^2+\sigma ^2}\bigg)-\bigg(\sigma ^2 \big(\alpha  L \bigg(\frac{1}{\big((\eta
		+L)^2+\sigma ^2\big)^{3/2}}+\frac{1}{\big((L-\eta )^2+\sigma ^2\big)^{3/2}}\bigg)\\&&+ \ \frac{\alpha
	}{\sqrt{(L-\eta )^2+\sigma ^2} \bigg(-\eta +\sqrt{(L-\eta )^2+\sigma ^2}+L\bigg)}+\frac{\alpha }{\sqrt{(\eta
			+L)^2+\sigma ^2} \bigg(\eta +\sqrt{(\eta +L)^2+\sigma ^2}+L\bigg)} \\&&+ \ 2\big)
			-2 \alpha \bigg)^2 \ . 
\end{eqnarray*}

\section{Poincar\'{e} section to string motion for general windings} \label{app A}

\begin{figure}[ H]
	\centering
	\hfill
	\begin{subfigure}{0.35\textwidth}
		\includegraphics[width=\textwidth]{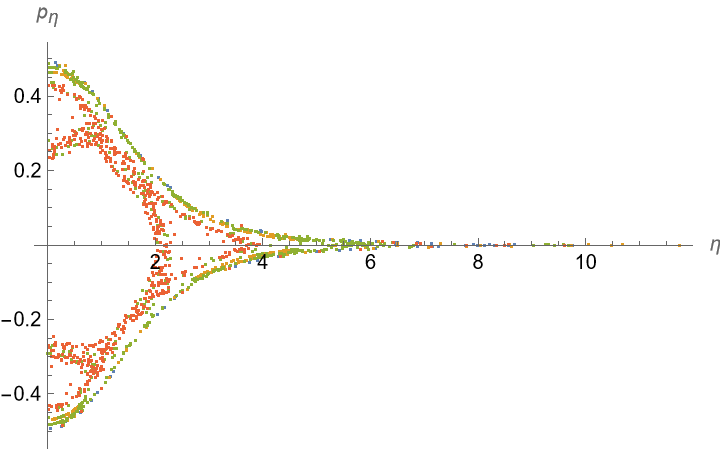}
		\caption{$E=35,  \ L=1,  \ k_1 = 2= k_2,  \ \alpha =0.1$. }
		\label{A1}
	\end{subfigure}
	\hfill
	\begin{subfigure}{0.35\textwidth}
		\includegraphics[width=\textwidth]{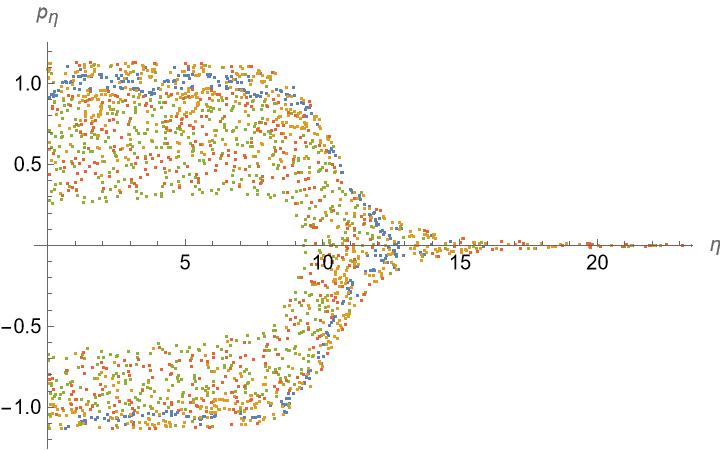}
		\caption{$E=35,  \ L=10,  \ k_1 = 2= k_2,  \ \alpha =0.1$.}
		\label{A2}
	\end{subfigure}
	\caption{Poincar\'{e} sections for winding $k_1 = k_2 =2$ with $L : \{1,10\}$ . Here we set the energy $E=35$ and $\alpha = 0.1$ in both cases.  }
\end{figure}

In this appendix, we study phase space plots corresponding to the string configurations with general values of the winding numbers $\{k_1, k_2\} >1$. 
For examples, in Figs.(\ref{A1})-(\ref{A2}) we consider the winding numbers $\{k_1 = k_2 = 2 \}$ and the parameter  $ L= \{1, 10\}$. 
It is very straightforward to see from Figs.(\ref{A1})-(\ref{A2}) that irrespective of the values of the winding numbers $k_i  (i=1,2)$ the Poincar\'{e} sections do not change qualitatively and exhibits chaotic motion in nature.

\section{Poincar\'{e} section to string motion in the non-Abelian T-dual background} \label{app B}

\begin{figure}[ H]
	\centering
	\hfill
	\begin{subfigure}{0.35\textwidth}
		\includegraphics[width=\textwidth]{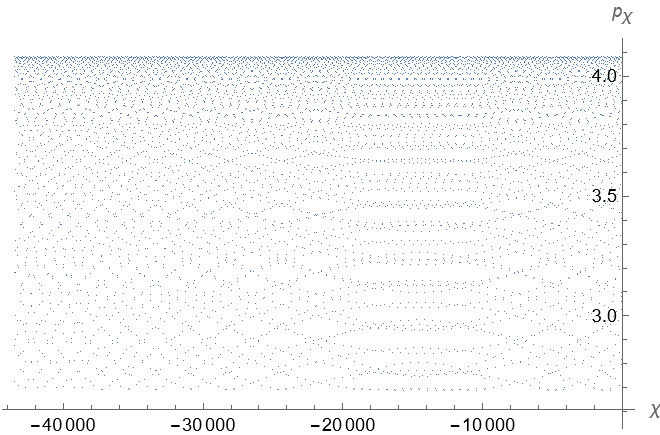}
		\caption{$k_1 = k_2 = 1$ .}
		\label{B1}
	\end{subfigure}
	\hfill
	\begin{subfigure}{0.35\textwidth}
		\includegraphics[width=\textwidth]{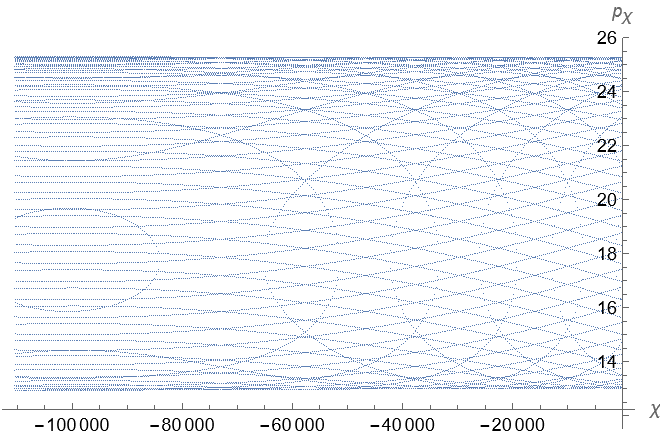}
		\caption{$k_1 = k_2 = 5$ .}
		\label{B2}
	\end{subfigure}
	\caption{Poincar\'{e} sections for non-Abelian T-dual solution. Here we set the energy $E=0.5$ \ ,  $\eta =1 \ , \ p_\eta = 0$ in both cases. }
\end{figure}

In the following we provide phase space plots corresponding to the string configurations moving in non-Abelian T-dual of AdS background. 
The corresponding potential function $V(\sigma, \eta)$ \cite{Lozano:2017ole} is given by
\begin{eqnarray} \label{pot nat}
 V(\sigma , \eta) = \eta \Bigg[\frac{ \sigma^2 }{2}- \ln \sigma -\frac{\eta^2}{3} \Bigg] \  . 
 \end{eqnarray} 
In the plots (Fig. (\ref{B1})-(\ref{B2})) we choose the 
winding numbers $\{k_1 = k_2 = 1\}$ and $\{k_1 = k_2 = 5\}$. 
Here the phase space dynamics exhibits integrable structure. 

\section{The Kovacic's algorithm} \label{app D}

The Kovacic algorithm \cite{K1,B1,K2} presents a systematic procedure for determining the integrability in the Liouvillian sense of a second-order linear homogeneous differential equation, described by:

\begin{equation}
	\label{Kov:Dif}
	\eta''(z)+M(z)\eta'(z)+N(z)\eta(z)=0 \, ,
\end{equation}
where $M(z)$, $N(z)$ are polynomial coefficients. Integrability in the Liouvillian sense signifies that solutions to this equation can be expressed using elementary functions, including algebraic, trigonometric, and exponential functions \cite{Roychowdhury:2017vdo,Nunez:2018ags,Nunez:2018qcj}.

Here, we will only cover the essential aspects of formalism, as the detailed mathematical analysis is complex. The objective is to determine the relationship among $M(z)$, $M'(z)$, and $N(z)$ that renders the differential equation (\ref{Kov:Dif}) integrable. To accomplish this, we initiate with a change of variable \cite{Roychowdhury:2017vdo,Nunez:2018ags,Nunez:2018qcj} of the following form:

\begin{equation}\label{Kov:Var}
	\eta(z) = \exp \bigg[\int d z \bigg(w(z)-\frac{M(z)}{2}\bigg)\,\bigg] .
\end{equation}

This change helps us rewrite the equation in a Schrödinger form \cite{Roychowdhury:2017vdo,Nunez:2018ags,Nunez:2018qcj}:

\begin{equation}\label{Kov:Fnl}
	w'(z) +w^{2}(z) =\mathcal{V}(z)=\frac{2M'(z)+M^{2}(z)-4N(z)}{4} \, .
\end{equation}
There will \textit{some necessary but not-sufficient conditions} on the combination of the functions $M(z)$, $M'(z)$, and $N(z)$ will make a different function called the potential function $\mathcal{V}(z)$ for which \eqref{Kov:Dif} to be \textit{Liouville integrable}. It has been shown that when the function $w(z)$ is algebraic of degrees 1, 2, 4, 6, or 12, equation \eqref{Kov:Dif} becomes Liouville integrable. This conclusion arises from applying Galois theory to differential equations, a method also known as Picard-Vessiot theory.

This formalism investigates the most general group of symmetries for the differential equation (\ref{Kov:Dif}). These symmetries refer to transformations that influence the solutions of the equation, constituting a subgroup of $SL(2,\mathbb{C})$. Kovacic identified four possible cases of subgroups within $SL(2,\mathbb{C})$. The symmetry transformations group $\mathcal{G}$, pertaining to solutions of the (\ref{Kov:Dif}), forms a subgroup of $SL(2,\mathbb{C})$: $\mathcal{G}\subset SL(2,\mathbb{C})$. Four notable cases, as discussed in \cite{Roychowdhury:2017vdo,Nunez:2018ags,Nunez:2018qcj}, include:

\begin{enumerate}
	\item If matrix of the following form generate the subgroup:
	\[
	\mathcal{G} = \begin{pmatrix} a & 0 \\ b & 1/a \end{pmatrix}, \quad a, b \in \mathbb{C}.
	\]
	For this case, $w(z)$ represents a rational function of degree $1$.
	
	\item If matrices of the following form generate the subgroup:
	\[
	\mathcal{G} = \begin{pmatrix} c & 0 \\ 0 & 1/c \end{pmatrix}, \quad
	\mathcal{G} = \begin{pmatrix} 0 & c \\ -1/c & 0 \end{pmatrix}, \quad c \in \mathbb{C}.
	\]
	Here, $w(z)$ signifies a rational function of degree $2$.
	
	\item $\mathcal{G}$ forms a finite group, excluding the aforementioned possibilities. In this case, $w(z)$ constitutes a rational function of degree either $4$, $6$, or $12$.
	
	\item The group $\mathcal{G}$ equates to $SL(2,\mathbb{C})$. If the solution $w(z)$ indeed exists, they are non-Liouvillian.
\end{enumerate}

It's worth noting that Kovacic not only developed an algorithm to find the solutions in the above first three cases, but also \textit{some necessary but not-sufficient conditions} on the potential function $\mathcal{V}(z)$ in \eqref{Kov:Fnl}. The $\mathcal{V}(z)$ must satisfy conditions for any of the first three cases mentioned above. The conditions are:


\begin{enumerate}
	
	\item	every pole of
	$\mathcal{V}(z)$  of order $1$ or $2n$ ($n \in \mathbb{Z}^{+}$). Moreover, the order of $\mathcal{V}(z)$ at infinity is either $2n$ or exceeds $2$.
	
	\item $\mathcal{V}(z)$ possesses either a single pole of order $2$, or poles of order $2n + 1$ greater than $2$.

	\item The poles of $V(z)$ do not exceed order $2$, and the order of $V(z)$ at infinity should be at least $2$.

	If none of the aforementioned conditions are satisfied, the analytic solution of (\ref{Kov:Dif}) (if it exists), is non-Liouvillian.

\end{enumerate}

\section{Brief review of type-IIA supergravity}

For massive type-IIA supergravity, the field strengths are given by
\begin{eqnarray}
H_3=dB_2 \ ; \ F_2= dC_1+F_0 B_2 \ ;  \ F_4=dC_3-H_3\wedge C_1 + \frac{F_0}{2} B_2\wedge B_2 \ .
\end{eqnarray}
The field strenghts are invariant under the gauge transformations
\begin{eqnarray}
\delta B_2= d\Lambda_1 \ ; \ \delta C_1= -F_0 \Lambda_1 \ ; \ \delta C_3= -F_0 \Lambda_1 \wedge B_2 \ ,
\end{eqnarray}
where $\Lambda$ is a one-form. 
The Bianchi identities become
\begin{eqnarray}
dH_3=0 \ ; \ dF_2=F_0H_3 \ ; \ dF_4=H_3\wedge F_2 \ .
\end{eqnarray}
The action of the massive type-IIA supergravity is 
\begin{eqnarray}
S =\frac{1}{2k^2} \int_{M_{10}} \sqrt{-g} &\Bigg[&e^{-2\Phi} \Big(R + 4(\partial\Phi)^2 -\frac{H_3^2}{12} \Big) - \frac{1}{2} \Big(F_0^2 + \frac{F_2^2}{2} +  \frac{F_4^2}{4!}\Big)\Bigg] \cr
&&-\frac{1}{2} \Bigg[dC_3 \wedge dC_3 \wedge B_2 + \frac{F_0}{3} dC_3 \wedge B^3 + \frac{F_0^2}{20} B_2^5 \Bigg] \ .
\end{eqnarray}
Einstein's equations are 
\begin{eqnarray}
R_{\mu\nu} + 2D_{\mu}D_{\nu}\Phi = \frac{1}{4} H_{\mu\nu}^2 + e^{2\Phi} \Bigg[\frac{1}{2} (F_2^2)_{\mu\nu} + \frac{1}{12} (F_4^2)_{\mu\nu} - \frac{1}{4} g_{\mu\nu} \Big(F_{0}^2 + \frac{1}{2} F_{2}^2 + \frac{1}{4!}F_{4}^2 \Big)\Bigg] \ .
\end{eqnarray}
The equation coming from varying the dilaton is 
\begin{eqnarray}
R + 4D^2\Phi - 4(\partial\Phi)^2 - \frac{1}{12}H_3^2=0 \ .
\end{eqnarray}
The equation of motion for the gauge fields in type-IIA case are given by  
\begin{eqnarray}
d\Big(e^{-2\Phi} \star H_3\Big) - F_2 \wedge \star F_4 - \frac{1}{2}  F_4 \wedge F_4 &=& F_0 \star F_2 \ , \cr
d\star F_2 + H_3 \wedge \star F_4 &=& 0 \ , \cr
d\star F_4 + H_3 \wedge F_4 &=& 0 \ .
\end{eqnarray}

\end{document}